\documentclass[fleqn,twoside]{article}
\usepackage{espcrc2}

\usepackage{graphicx}
\usepackage[figuresright]{rotating}

\newcommand{\AmS}{{\protect\the\textfont2
  A\kern-.1667em\lower.5ex\hbox{M}\kern-.125emS}}

\title{\bf Charged LFV in a low-scale seesaw mSUGRA model}

\author{Amon Ilakovac\address
       {University of Zagreb, Department of Physics, Bijeni\v cka
        cesta 32, P.O.Box 331, Zagreb, Croatia} and
        Apostolos Pilaftsis\address{School of Physics and Astronomy,
          Univ.~of Manchester, 
        Manchester M13 9PL, United Kingdom}}
       
\begin{document}

\begin{abstract}
We  investigate the influence  of the  boundary conditions  of minimal
supergravity  (mSUGRA)  on  the  supersymmetric mechanism  for  lepton
flavour violation (LFV)  proposed recently~\cite{PRD80_PI}, within the
framework of  the MSSM extended by TeV-scale  singlet heavy neutrinos.
We find  that the consideration  of the mSUGRA boundary  condition may
increase  the branching  ratios of  the muon  and tauon  decaying into
three  charged  leptons   by  up  to  a  factor   of  5,  whereas  the
corresponding branching ratio for their photonic decays remains almost
unchanged.
\vspace{1pc}
\end{abstract}

\maketitle

\section{SLFV in the MSSM3N}

Recently,  we proposed~\cite{PRD80_PI}  a novel,  fully supersymmetric
mechanism   for    LFV,   which    is   independent   of    the   soft
supersymmetry~(SUSY) breaking sector of  the theory. The mechanism was
called supersymmetric lepton flavour violation (SLFV).  To demonstrate
the importance  of SLFV, we considered a  $R$-parity conserving seesaw
extension  of the  MSSM with  one  singlet heavy  neutrino per  family
(MSSM3N).   The leptonic sector  of the  superpotential of  the MSSM3N
reads~\cite{PRD80_PI}:
\begin{equation}
  \label{Wpot}
W_{\rm l}\, =\, \widehat{E}^C {\bf h}_e \widehat{H}_d
\widehat{L} + \widehat{N}^C {\bf h}_\nu \widehat{L} \widehat{H}_u+
\widehat{N}^C {\bf m}_M \widehat{N}^C 
\end{equation}
where the complex $3\times 3$ matrices, ${\bf h}_e$, ${\bf h}_\nu$ and
${\bf m}_M$ represent the  electron and the neutrino Yukawa couplings,
and the  symmetric heavy neutrino Majorana  mass matrix, respectively.
We  assume that  the heavy  neutrino sector  of the  model  is $SO(3)$
symmetric being broken down to an $U(1)$ lepton symmetry by the Yukawa
sector~\cite{AP_ResLep}. The approximate  breaking of these symmetries
lead to almost degenerate heavy neutrinos, $m_M\approx m_N {\bf 1}_3$.
The approximate  flavour symmetries  also assure small  light neutrino
masses, permitting heavy neutrino mass scale as low as $100$~GeV, in a
way such that  the usual see-saw suppression factor  of LFV processes,
$m_\nu/m_N$ is avoided.  The SLFV effects depend on the LFV parameters
\begin{equation}
  \label{Omega}
\mbox{\boldmath{$\Omega$}}_{\ell\ell'} \; =\;  \frac{v^2_u}{2
  m^2_N}\ ({\bf  h}^\dagger_\nu 
{\bf h}_\nu)_{ll'}\ .
\end{equation}
In contrast to usual SUSY studies \cite{usualSUSY}  
where LFV effects only depend on the
flavour  structure  of  the   soft  SUSY-breaking  sector  induced  by
renormalization group  (RG) running, SLFV  effects only depend  on the
superpotential heavy neutrino mass scale $m_N$ and the neutrino-Yukawa
couplings  ${\bf  h}^\nu$.   In  addition,  they  depend  on  $\langle
\sqrt{2} H_u\rangle \equiv v_u  = v\cos\beta$, with $v\approx 246$~GeV
and $\tan\beta = \langle H_u\rangle/\langle H_d\rangle$.

The  SLFV  amplitudes  are   induced  by  heavy  neutrinos  and  heavy
sneutrinos at  the one-loop  level.  The loop  sneutrino contributions
include  heavy sleptons and/or  heavy squarks,  and therefore  they do
depend   on    the   soft   SUSY-breaking    sparticle   masses.    In
Ref.~\cite{PRD80_PI},  the   LFV  observables  induced   by  SLFV  are
evaluated by selecting typical values  for the sparticle masses at the
electroweak scale.   Here, we extend  the previous study  and evaluate
the soft SUSY-breaking parameters,  as functions of the heavy neutrino
mass scale  $m_N$ and  the LFV parameters  $\Omega_{\ell\ell'}$, using
one-loop MSSM3N RG equations~\cite{RGE_MSSM3N} with universal boundary
conditions  at the  gauge-coupling  unification scale  $M_X =  2.5\times
10^{16}$~GeV.  RG analysis confirms  that the heavy neutrino sector is
supersymmetric  almost for the  whole parameter  space allowed  by the
perturbative condition on neutrino  Yukawa matrices: ${\rm Tr}\, ({\bf
  h}_\nu^\dagger    {\bf    h}_\nu)    <4\pi$.    Specifically,    the
singlet-neutrino  sector is  supersymmetric to  a  good approximation,
provided the  heavy neutrino  mass $m_N$ is  of comparable order or
larger than the soft SUSY breaking parameters $m_0$ (scalar mass), $M$
(gaugino  mass) and  $A_0$ (trilinear  scalar coupling).  In  the same
kinematic   regime,  the   light  left-handed   sneutrinos   are  also
degenerate.   The  superpotential  $\widehat{H}_u\widehat{H}_d$-mixing
term $\mu$ turns out to be typically of order $400$~GeV, and therefore
smaller than the heavy neutrino mass scale $m_N$, for the largest part
of   the   allowed  parameter   space.    The   above  justifies   the
approximations  used  to  obtain   the  dominant  terms  of  the  SLFV
amplitudes in Ref.~\cite{PRD80_PI}.

Within  the  above  framework,  we  may  calculate  the  leading  SLFV
amplitudes close  to the SUSY limit  in the lowest order  of $v_u$ and
$m_N^{-1}$.   To  leading order  in  $g_W$  and  ${\bf h}_{\nu}$,  the
pertinent LFV amplitudes read~\cite{PRD80_PI}:
\begin{eqnarray}
  \label{Trans}
{\cal T}^{\gamma l'l}_\mu 
 \!\!&=&\!\! 
\frac{e\, \alpha_w}{8\pi M^2_W}\
\bar{l}' \Big( F_\gamma^{l'l}\, q^2 \gamma_\mu P_L 
\nonumber\\
 \!\!&+&\!\!
G^{l'l}_\gamma\,
i\sigma_{\mu\nu} q^\nu m_l P_R \Big) l\;,\nonumber\\
{\cal T}^{Z l'l}_\mu \!\!&=&\!\!  \frac{g_w\, \alpha_w}{8\pi \cos\theta_w}\
F^{l'l}_Z\, \bar{l}' \gamma_\mu P_L l\; ,\\
{\cal T}^{l'l_1l_2}_l \!\!\!&=&\!\!  -\frac{\alpha^2_w}{4 M^2_W}\;
F^{ll'l_1l_2}_{\rm box}\, \bar{l}'\gamma_\mu P_L l\;
\bar{l}_1\gamma^\mu P_L l_2 \; ,\nonumber
\end{eqnarray}
where  $q  =  p_{\ell'}   -  p_\ell$.   The  amplitudes  {\small  ${\cal
    T}^{l'u_1u_2}_l$}  and {\small  ${\cal T}^{l'd_1d_2}_l$}  have the
same structure  as the amplitude {\small  ${\cal T}^{l'l_1l_2}_l$}, up
to replacements  $\ell_i \to u_i  \to d_i,\ i=1,2$.  The  form factors
{\small   $F^{l'l}_\gamma$},    {\small   $G^{l'l}_\gamma$},   {\small
  $F^{l'l}_Z$},    {\small    $F^{ll'l_1l_2}_{\rm   box}$},    {\small
  $F^{ll'u_1u_2}_{\rm  box}$} and  {\small  $F^{ll'd_1d_2}_{\rm box}$}
receive contributions  from both  the heavy neutrinos  $N_{1,2,3}$ and
the right-handed sneutrinos $\widetilde{N}_{1,2,3}$. To illustrate the
importance  of the  SLFV  effects, we  give  the leading  form of  the
form factors    {\small    $F^{l'l}_\gamma$},   {\small $G^{l'l}_\gamma$},    
{\small $F^{l'l}_Z$}, {\small $F^{ll'l_1l_2}_{\rm box}$}, 
{\small $F^{ll'u_1u_2}_{\rm box}$} and {\small $F^{ll'd_1d_2}_{\rm box}$}
in the Feynman gauge,
\begin{eqnarray}
  \label{Fgamma}
(F^{l'l}_\gamma)^N
 \!\!\!\!\!&=&\!\!\!\!\!
 \frac{ \mbox{\boldmath{$\Omega$}}_{\ell\ell'} }{6\,s^2_\beta}\,
  \ln \lambda_N\; ,
\nonumber\\
(F^{l'l}_\gamma)^{\widetilde{N}}
 \!\!\!\!\!&=&\!\!\!\!\!
  \frac{\mbox{\boldmath{$\Omega$}}_{\ell\ell'}}{3\,s^2_\beta}\,
  \sum_{k=1}^2 {\cal V}_{k2}^2
  \ln \lambda_{Nk}\; ,
\\
  \label{Ggamma}
(G^{l'l}_\gamma)^N
 \!\!\!\!\!&=&\!\!\!\!\!
  \mbox{\boldmath{$\Omega$}}_{\ell\ell'}\,
  \bigg( -\frac{1}{6\,s^2_\beta} - \frac{5}{6}\, \bigg)\;,
\nonumber\\
(G^{l'l}_\gamma)^{\widetilde{N}} 
 \!\!\!\!\!&=&\!\!\!\!\!
\mbox{\boldmath{$\Omega$}}_{\ell\ell'}\, \Bigg(\, \frac{1}{6\,s^2_\beta}  +
  f\,\bigg)
\; ,
\\
  \label{FZ}
(F^{l'l}_Z)^N
 \!\!\!\!\!&=&\!\!\!\!\!
 -\, \frac{3\, \mbox{\boldmath{$\Omega$}}_{\ell\ell'}}{2}\,
  \ln \lambda_N - \frac{( \mbox{\boldmath{$\Omega$}}_{\ell\ell'}^2) }{2\,s^2_\beta} \,
  \lambda_N\; , 
\nonumber\\
(F^{l'l}_Z)^{\widetilde{N}}
 \!\!\!\!\!&=&\!\!\!\!\!
 \mbox{\boldmath{$\Omega$}}_{\ell\ell'}\, g\, \ln \lambda_N
\\
F^{ll'l_1l_2}_{\rm box}
 \!\!\!\!\!&=&\!\!\!\!\!
 -\, (\delta_{\ell_1\ell_2}\Omega_{\ell\ell'} + \delta_{\ell'\ell_2}\Omega_{\ell\ell_1})
\nonumber\\
 \!\!\!\!\!&+&\!\!\!\!\!
 ( \Omega_{\ell\ell'} \Omega_{\ell_2\ell_1} 
 + \Omega_{\ell\ell_1} \Omega_{\ell_2\ell'} ) \frac{\lambda_N}{4 s_\beta^4}
\nonumber\\
F^{ll'l_1l_2}_{\rm box}
 \!\!\!\!\!&=&\!\!\!\!\!
 (\delta_{\ell_1\ell_2}\Omega_{\ell\ell'} + \delta_{\ell'\ell_2}\Omega_{\ell\ell_1})\, h_\ell
\nonumber\\
 \!\!\!\!\!&+&\!\!\!\!\!
 ( \Omega_{\ell\ell'} \Omega_{\ell_2\ell_1} 
 + \Omega_{\ell\ell_1} \Omega_{\ell_2\ell'} ) \frac{\lambda_N}{4 s_\beta^4}
\\
F^{ll'u_1u_2}_{\rm box}
 \!\!\!\!\!&=&\!\!\!\!\!
 4 \Omega_{\ell\ell'}
\nonumber\\
F^{ll'u_1u_2}_{\rm box}
 \!\!\!\!\!&=&\!\!\!\!\!
 \Omega_{\ell\ell'} h_u
\\
F^{ll'd_1d_2}_{\rm box}
 \!\!\!\!\!&=&\!\!\!\!\!
 - \Omega_{\ell\ell'}
\nonumber\\
F^{ll'd_1d_2}_{\rm box}
 \!\!\!\!\!&=&\!\!\!\!\!
 \Omega_{\ell\ell'} h_d 
\end{eqnarray}
where    {\small   $\lambda_N    =    \frac{m_N^2}{M_W^2}$},   {\small
  $\lambda_{Nk} =  \frac{m_N^2}{m^2_{\tilde{\chi}_k}}$}, ${\cal V}$ is
one of the unitary matrices diagonalizing the chargino mass matrix and
$f$,  $g$, $h_\ell$,  $h_u$  and $h_d$  are  complicated functions  of
masses and  mixing matrices.  Detailed results of  this study  will be
given in a forthcoming publication \cite{IPP}.

In    the    SUSY   limit    $\tan\beta    \to    1$,   $\mu\to    0$,
$m_{\tilde{\chi}_k}\to M_W$, $f\to\frac{5}{6}$, $g\to \frac{3}{2}$,
$h_\ell\to -1$, $h_u\to 0$ and $h_d\to -1$.  We
note that  the photonic dipole  form factor {\small  $G^{l'l}_\gamma =
  (G^{l'l}_\gamma)^N +  (G^{l'l}_\gamma)^{\widetilde{N}}$} vanishes in
the SUSY limit. This is  a consequence of the SUSY non-renormalization
theorem~\cite{FR}.  Beyond the SUSY  limit, it strongly depends on the
soft SUSY-breaking sector and particulary on the sparticle masses.

In  all   formfactors,  except  of  $G^{l'l}_\gamma$,   the  $N$-  and
$\widetilde{N}$-loop contributions  add constructively.  Specifically,
in the $M_{SUSY}\gg  M_W$ and in the large  $M_N$ limit, the following
approximate    form    factor    relations    are    valid:    {\small
  $F_\gamma^{\ell\ell'} \approx  3 (F_\gamma^{\ell\ell'})^N$}, {\small
  $| G_\gamma^{\ell\ell'}|                           \stackrel{<}{{}_\sim}
  |(G_\gamma^{\ell\ell'})^N|$},    {\small    $F_Z^{\ell\ell'}   \approx
  (F_Z^{\ell\ell'})^N$},  {\small   $F^{ll'l_1l_2}_{\rm  box}  \approx
  (F^{ll'l_1l_2}_{\rm box})^N$}, {\small $F^{ll'd_1d_2}_{\rm box} \le 2
    (F^{ll'd_1d_2}_{\rm box})^N$}, and {\small $F^{ll'u_1u_2}_{\rm box}
      \approx (F^{ll'u_1u_2}_{\rm box})^N$}.   It is important to note
    that the large $m_N$ limit corresponds to a kinematic region where
    the  neutrino Yukawa couplings  ${\bf h}_\nu$  are large  (see Eq.
    (\ref{Omega})).  In this limit, the $\mbox{\boldmath{$\Omega$}}^2$
    terms  dominate   in  $Z$  and  leptonic   box  amplitudes.   More
    precisely,         the        terms         proportional        to
    $\mbox{\boldmath{$\Omega$}}^2$    become    larger   than    those
    proportional  to  $\mbox{\boldmath{$\Omega$}}$,  if  $g_w^2<  {\rm
      Tr}\, ({\bf h}^\dagger_\nu {\bf h}_\nu)$.

\begin{figure}[htp]
 \centering
 \includegraphics[clip,width=0.45\textwidth]{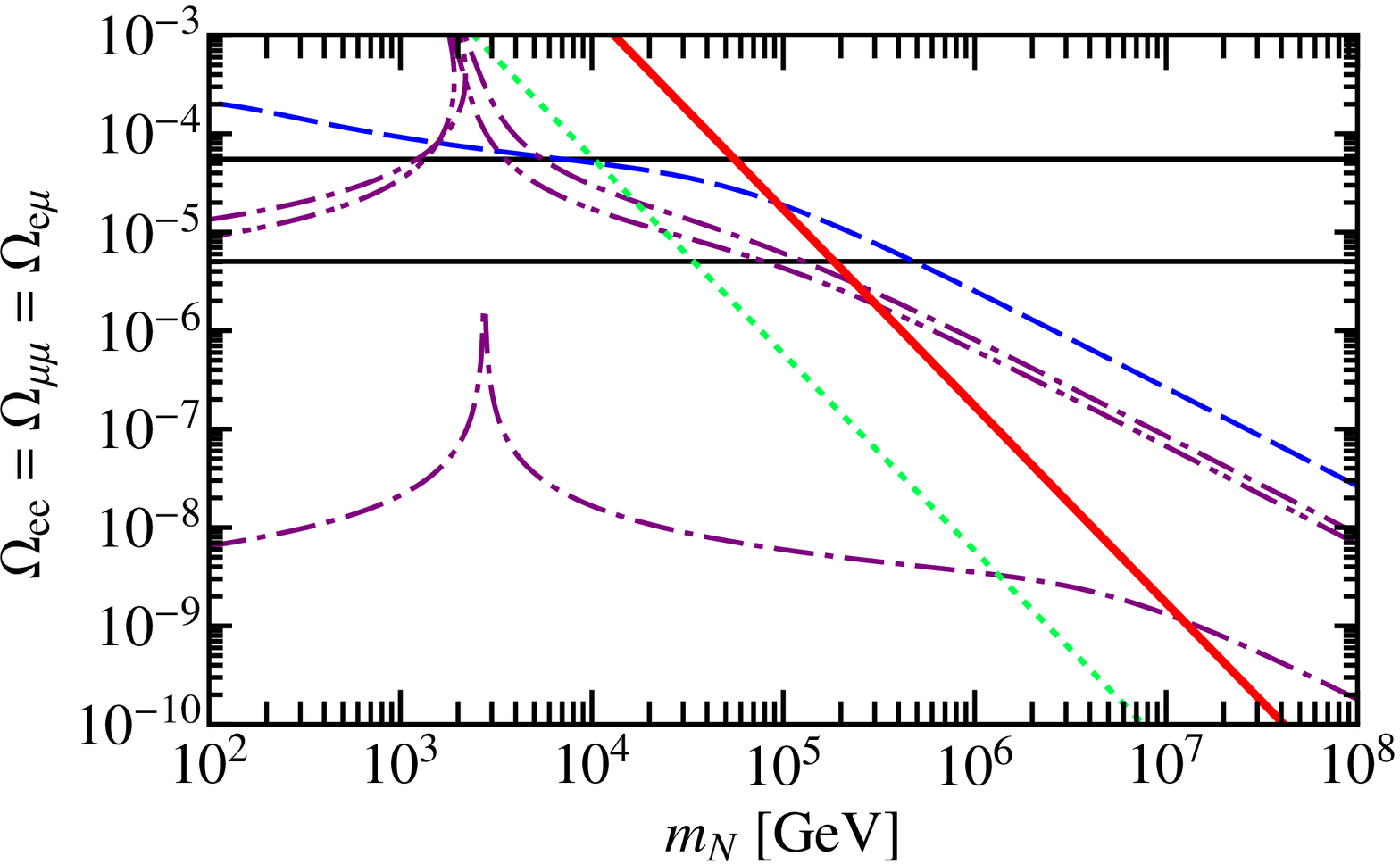}\\[3mm]
 \includegraphics[clip,width=0.45\textwidth]{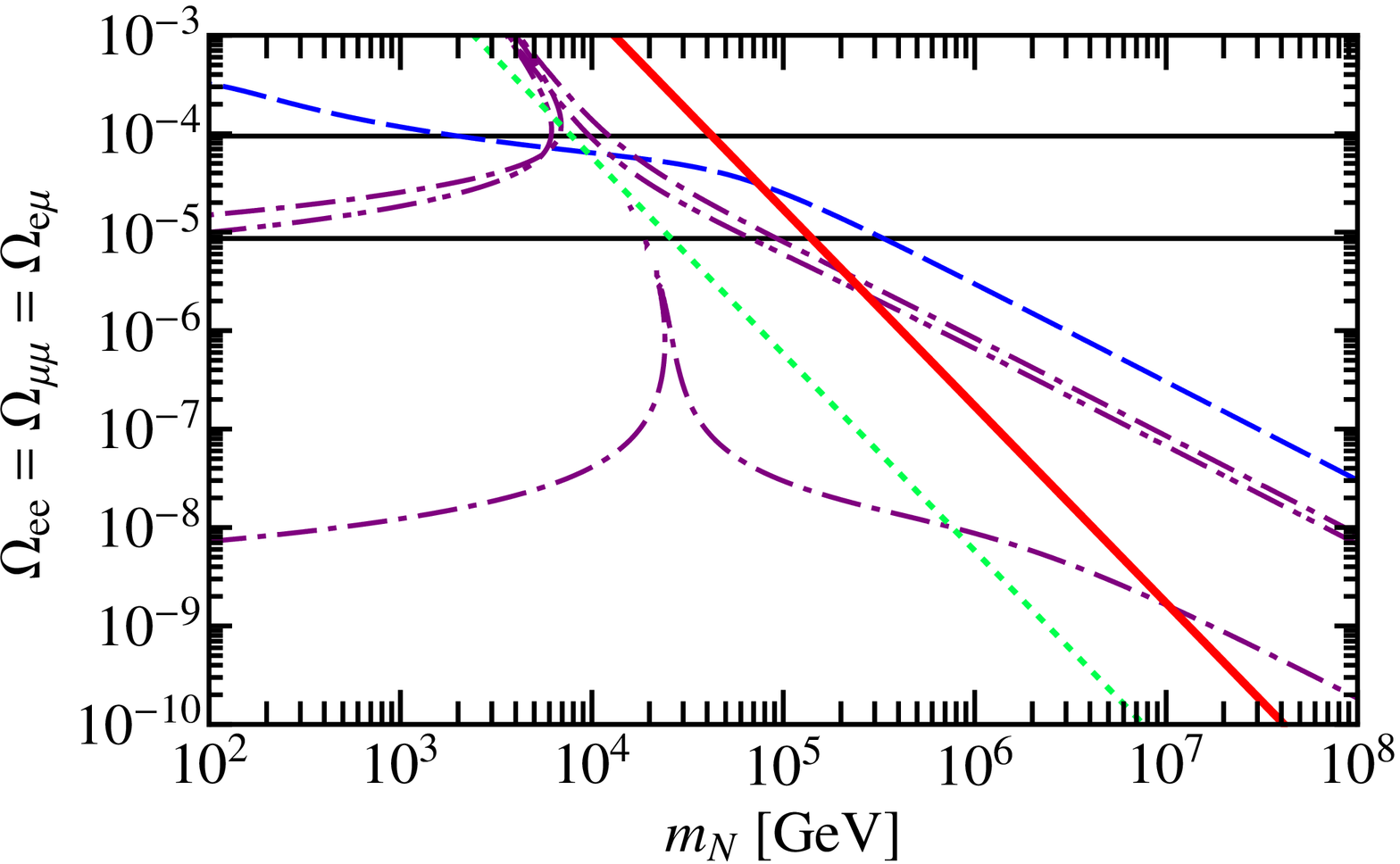}\\[3mm]
 \includegraphics[clip,width=0.45\textwidth]{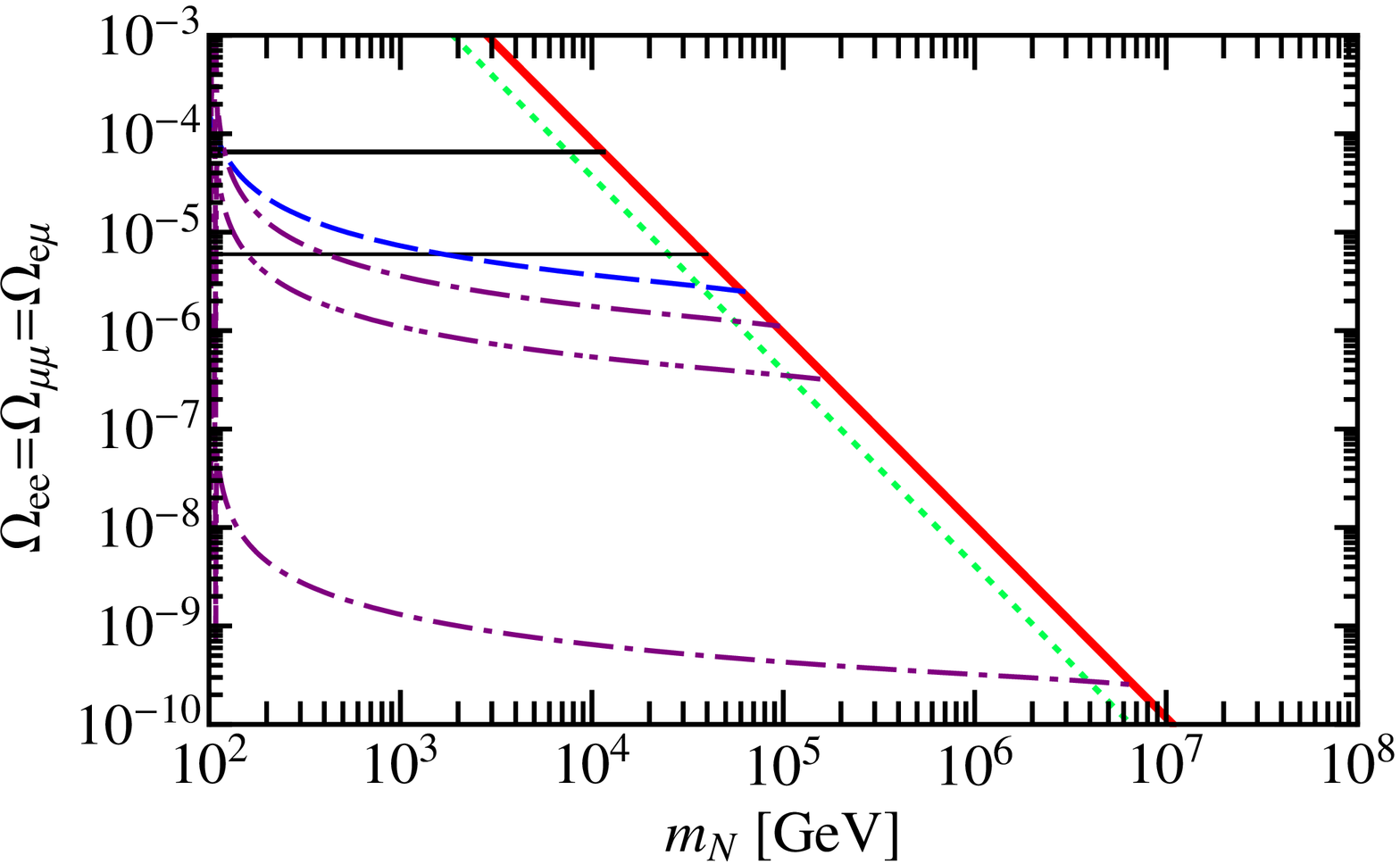}
\caption{Exclusion  contours  of  ${\bf \Omega}_{e\mu}$  versus  $m_N$
  derived  from  experimental   limits  on  $B(\mu^-  \to  e^-\gamma)$
  (solid), $B(\mu^- \to e^-e^-e^+)$ (dashed) and $\mu\to e$ conversion
  in  Titanium (dash-dotted)  and Gold  (dash-double-dotted), assuming
  ${\bf  \Omega}_{ee} = {\bf  \Omega}_{\mu\mu} =  {\bf \Omega}_{e\mu}$
  and other ${\bf \Omega}_{\ell\ell'} = 0$. The upper, middle and lower panel
  represent the  exclusion contours in  the SM3N, the MSSM3Nf  and the
  MSSM3NS, respectively.  The areas  above the contours  are excluded;
  see the text for more details.}
  \label{f1}  
\end{figure}


\begin{figure}[htp]
 \centering
  \includegraphics[clip,width=0.45\textwidth]{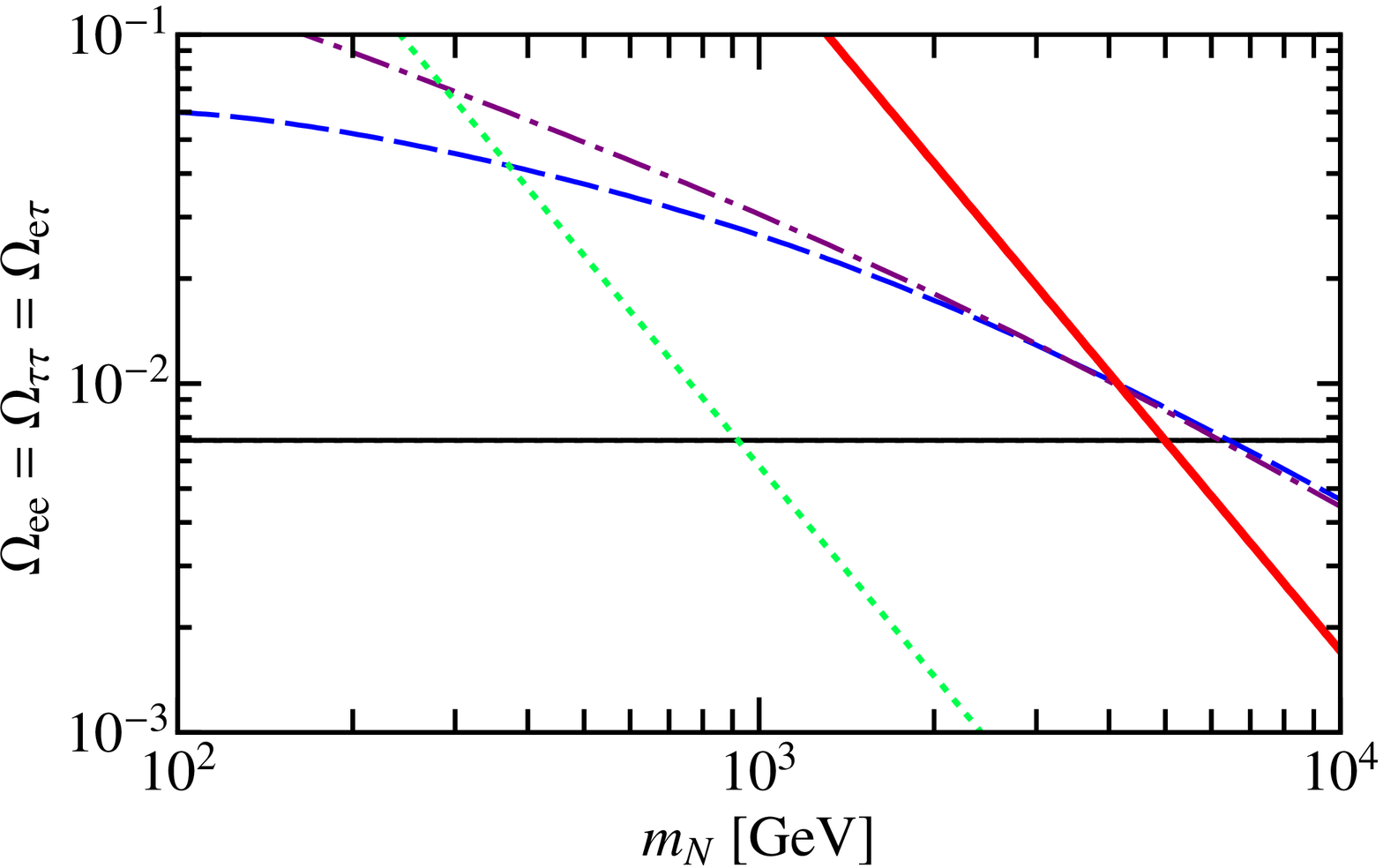}\\[3mm]
  \includegraphics[clip,width=0.45\textwidth]{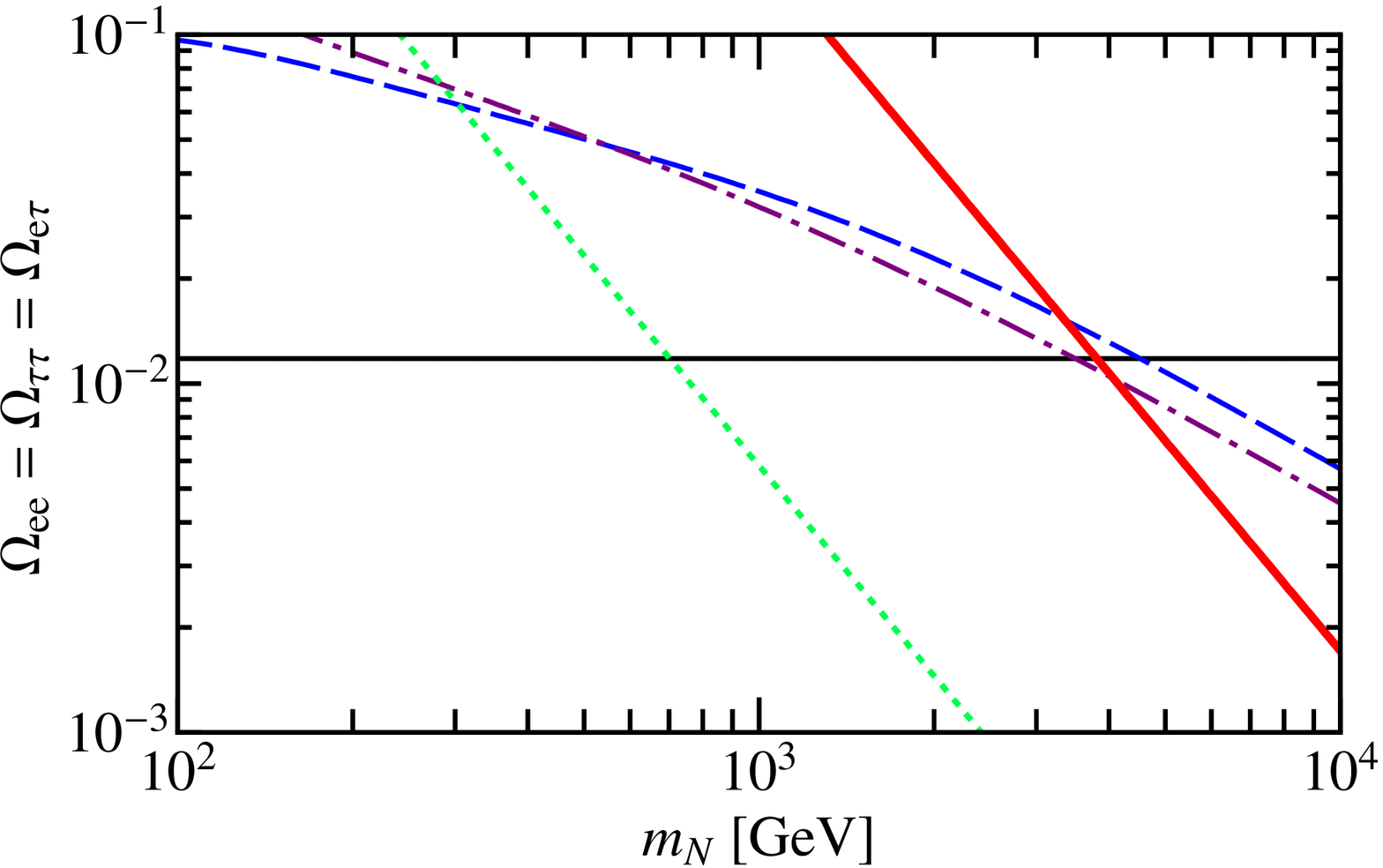}\\[3mm]
  \includegraphics[clip,width=0.45\textwidth]{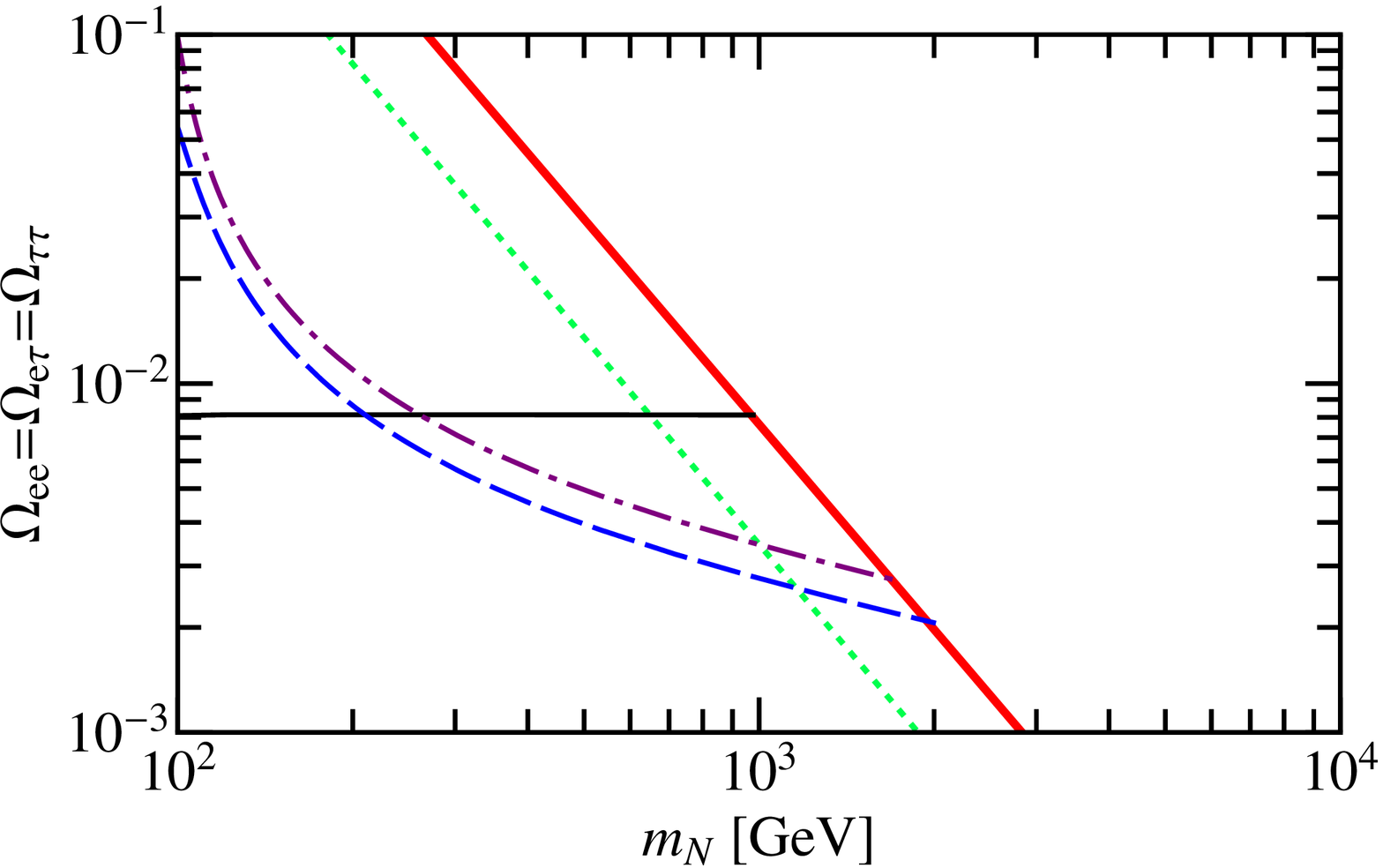}
\caption{Exclusion  contours of  ${\bf  \Omega}_{e\tau}$ versus  $m_N$
  derived  from present  experimental  upper limits  on $B(\tau^-  \to
  e^-\gamma)$   (solid),  $B(\tau^-   \to  e^-e^-e^+)$   (dashed)  and
  $B(\tau^-\to  e^-\mu^-\mu^+)$  (dash-dotted),  assuming  that  ${\bf
    \Omega}_{ee} = {\bf \Omega}_{\tau\tau} = {\bf \Omega}_{e\tau}$ and
  other ${\bf  \Omega}_{\ell\ell'} =  0$.  The upper,  middle  and lower  panel
  represent the  exclusion contours in  the SM3N, the MSSM3Nf  and the
  MSSM3NS, respectively.  The areas  above the contours  are excluded;
  more details are given in the text.}
  \label{f2}
\end{figure}

\begin{figure}[htp]
 \centering
  \includegraphics[clip,width=0.45\textwidth]{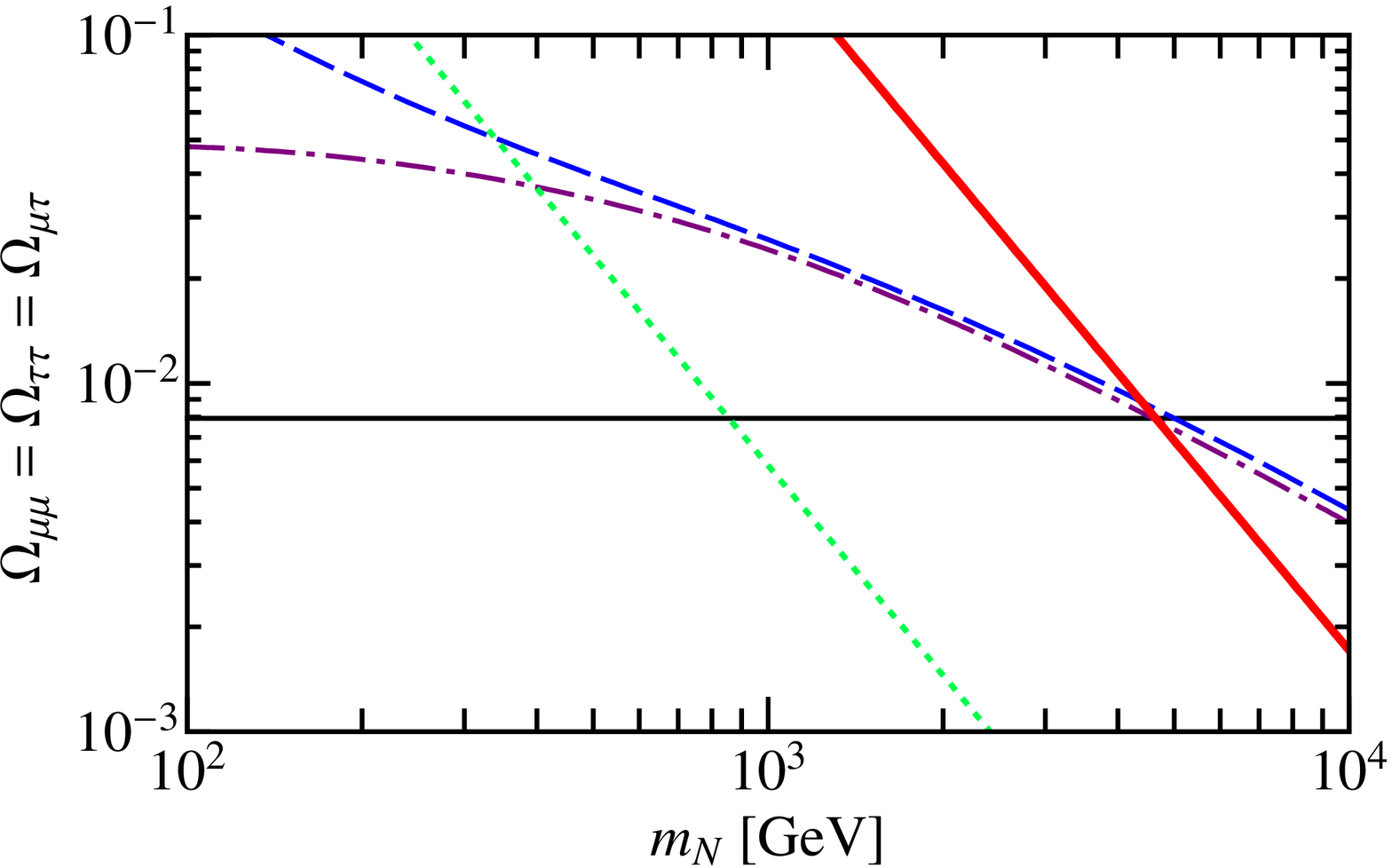}\\[3mm]
  \includegraphics[clip,width=0.45\textwidth]{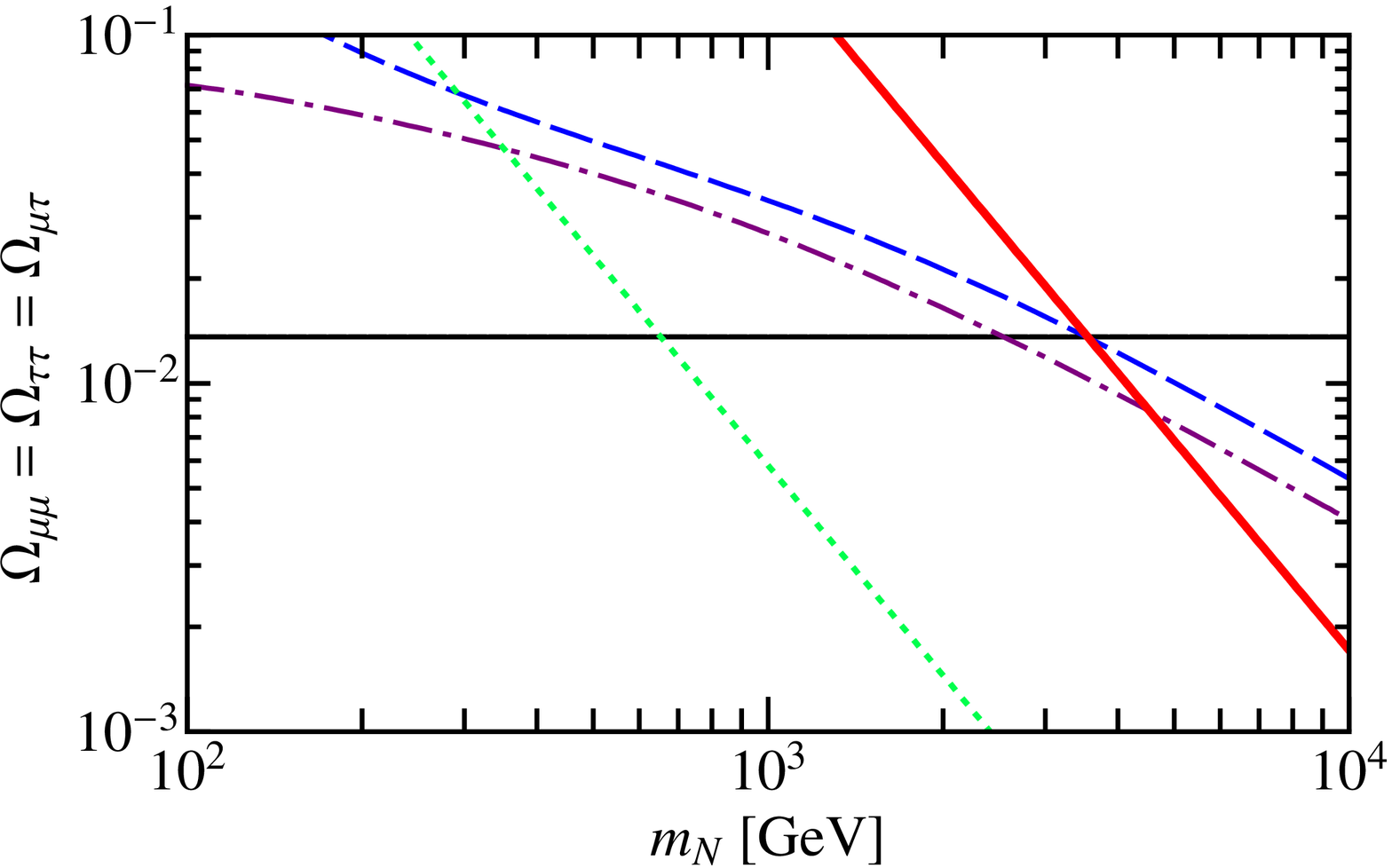}\\[3mm]
  \includegraphics[clip,width=0.45\textwidth]{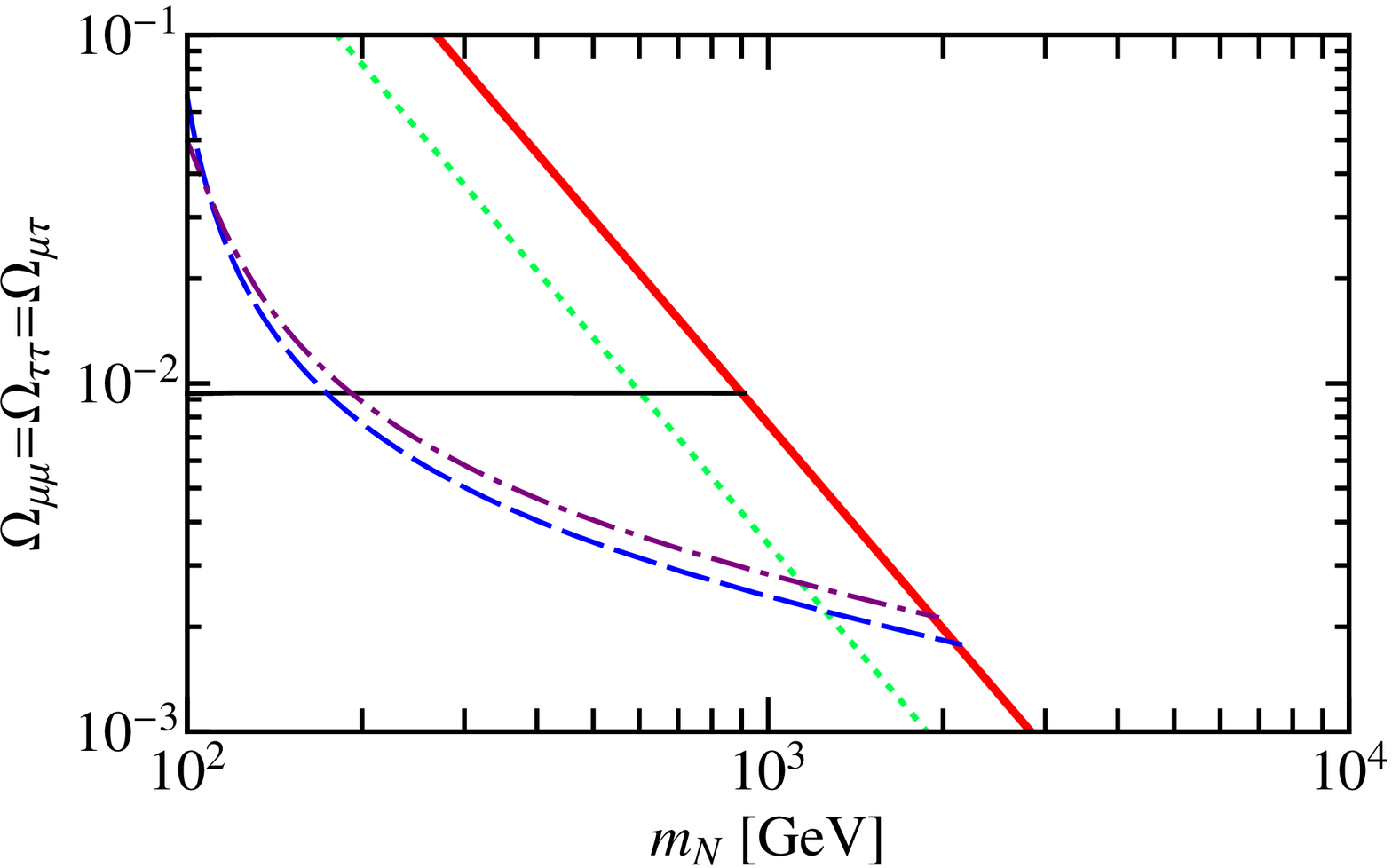}
\caption{Exclusion contours  of ${\bf \Omega}_{\mu\tau}$  versus $m_N$
  derived  from present  experimental  upper limits  on $B(\tau^-  \to
  \mu^-\gamma)$ (solid), $B(\tau^-  \to \mu^-\mu^-\mu^+)$ (dashed) and
  $B(\tau^-\to   \mu^-e^-e^+)$  (dash-dotted),  assuming   that  ${\bf
    \Omega}_{\mu\mu}     =    {\bf    \Omega}_{\tau\tau}     =    {\bf
    \Omega}_{\mu\tau}$ and other ${\bf \Omega}_{\ell\ell'}  = 0$. The upper, middle
  and lower  panel represent the  exclusion contours in the  SM3N, the
  MSSM3Nf and the MSSM3NS,  respectively. The areas above the contours
  are excluded; more details are given in the text.}
  \label{f3}
\end{figure}

\section{Numerical estimates}

We  now  present  numerical  estimates  of LFV  observables  in  three
distinct models:  (i) the Standard  Model with one  heavy right-handed
neutrino per  family~(SM3N); (ii)  the MSSM3N with  fixed superpartner
masses~(MSSM3Nf);    (iii)    the    MSSM   with    mSUGRA    boundary
conditions~(MSSM3NS).

In    the    SM3N,    the    LFV    amplitudes    depend    only    on
$\mbox{\boldmath{$\Omega$}}_{\ell\ell'}$  and  $m_N$.   For  the  SUSY
models, the MSSM3Nf and the  MSSM3NS, the LFV amplitudes are functions
of   $\mbox{\boldmath{$\Omega$}}_{\ell\ell'}$,   $m_N$,   $\tan\beta$,
$\mu$,   the   slepton  and/or   squark   masses,   and  the   unitary
chargino-mixing mass  matrices. We  take $\tan\beta =  3$, which  is a
value close to the SUSY limit  value $\tan\beta = 1$.  In the MSSM3Nf,
the  lepton  and  squark  matrices   and  $\mu$  are  taken  as  input
parameters.  We  fix  $-\mu  =  \widetilde{M}_Q  =  M_{\tilde{\nu}}  =
200$~GeV, $M_{\widetilde{W}}  = 100$~GeV and $\tan\beta =  3$.  In the
MSSM3NS, the $\mu$ parameter and the sparticle masses are functions of
$m_N$   and    $\mbox{\boldmath{$\Omega$}}_{\ell\ell'}$.    They   are
determined  by the MSSM3N  RGEs and  the universal  soft SUSY-breaking
parameters  $M$,  $m_0$  and   $A_0$  defined  at  the  gauge-coupling
unification  scale  $M_X$.   As  input values,  we  take  $M=250$~GeV,
$m_0=100$~GeV and $A_0=150$~GeV.

To simplify our analysis of identifying the regions of parameter space
excluded  by experimental  limits of  LFV processes  $\mu\to  e+X$ and
$\tau\to e + X$ (where $X$ indicates generically a photon or 3 charged
leptons), we consider three separate conservative scenarios with three
non-zero      $\mbox{\boldmath{$\Omega$}}_{\ell\ell'}$     parameters:
$\mbox{\boldmath{$\Omega$}}_{\mu\mu} = \mbox{\boldmath{$\Omega$}}_{\mu
  e}                =                \mbox{\boldmath{$\Omega$}}_{ee}$,
$\mbox{\boldmath{$\Omega$}}_{\tau\tau}                                =
\mbox{\boldmath{$\Omega$}}_{\tau                  e}                 =
\mbox{\boldmath{$\Omega$}}_{ee}$,                                   and
$\mbox{\boldmath{$\Omega$}}_{\tau\tau}                                =
\mbox{\boldmath{$\Omega$}}_{\tau\mu}                                  =
\mbox{\boldmath{$\Omega$}}_{\mu\mu}$ respectively.

The  exclusion  contours of  ${\bf  \Omega}_{e\mu}$  versus $m_N$  for
$\mu\to  e+X$ processes,  ${\bf  \Omega}_{e\tau}$ versus  $m_N$ for
$\tau\to   e  +   X$ and ${\bf  \Omega}_{\mu\tau}$ versus  $m_N$ for
$\tau\to   \mu  +   X$ are  given   in  Figs.~\ref{f1},~\ref{f2} and~\ref{f3},
respectively.   The  areas  above  the  curves are  forbidden  by  the
experimental upper  bounds on  the corresponding processes.   The area
above  the diagonal  solid lines  represent the  nonpertubative regime
with ${\rm Tr}\,({\bf h}^\dagger_\nu  {\bf h}_\nu) > 4\pi$, whilst the
area above the diagonal dotted  lines represent the region where the
Yukawa  couplings  dominate the  LFV  observables,  ${\rm Tr}\,  ({\bf
  h}^\dagger_\nu {\bf  h}_\nu) > g_w^2$.   The higher values  of ${\bf
  \Omega}_{\ell\ell'}$  correspond  to   smaller  values  of  the  LFV
observables:  the  factors   multiplying  the  combinations  of  ${\bf
  \Omega}_{\ell\ell'}$     elements    is     smaller,     if    ${\bf
  \Omega}_{\ell\ell'}$ needed to  satisfy the experimental upper bound
is larger.

The  mSUGRA   boundary  condition  has  a  strong   influence  on  the
perturbativity condition: ${\rm Tr}\,({\bf h}^\dagger_\nu {\bf h}_\nu)
> 4\pi$. In the SM3N and  the MSSM3Nf, the condition ${\rm Tr}\, ({\bf
  h}^\dagger_\nu  {\bf h}_\nu)  >  4\pi$ is  determined  at the  $M_Z$
scale. In the MSSM3NS, ${\rm Tr}\, ({\bf h}^\dagger_\nu {\bf h}_\nu) >
4\pi$ has to be satisfied for any RG scale between $M_Z$ and $M_X$. As
${\bf  h}_\nu$  increases with  the  RG  scale,  the ${\rm  Tr}\,({\bf
  h}^\dagger_\nu  {\bf h}_\nu)  >  4\pi$ is  determined  at the  gauge
unification  scale,  when  the  typical value  for  ${\rm  Tr}\,({\bf
 h}^\dagger_\nu {\bf h}_\nu)$ at the $M_Z$ scale is $\sim 0.3-0.45$. Thus,
signficant  part  of the  SM3N  and  MSSM3Nf  parameter space  in  the
$m_N$-${\bf \Omega}_{\ell\ell'}$  plane gets excluded  in the MSSM3NS.
Also, the boundary lines  ${\rm Tr}\,({\bf h}^\dagger_\nu {\bf h}_\nu)
= 4\pi$  and ${\rm  Tr}\, ({\bf h}^\dagger_\nu  {\bf h}_\nu)  = g_w^2$
come closer  to each other.   Moreover, the LFV observables  cannot be
evaluated   beyond  the   perturbativity  limit   ${\rm   Tr}\,  ({\bf
  h}^\dagger_\nu  {\bf  h}_\nu)  =   4\pi$,  since  the  RGEs  rapidly
diverge. Instead, in the SM3N  and MSSM3Nf, the LFV observables can be
computed for any value of $m_N$ and ${\bf \Omega}_{\ell\ell'}$.

Figures \ref{f1},  \ref{f2} and \ref{f3} contain 3  panels. The upper,
the middle and the lower panels display exclusion contours obtained in
the SM3N, the MSSM3Nf and the MSSM3NS, respectively.

Figure~\ref{f1} presents  exclusion contours for  current experimental
limits on  and future  sensitivities to LFV  processes of $\mu  \to e$
transitions: $B(\mu^- \to  e^-\gamma) < 1.2\times 10^{-11}$~\cite{PDG}
(upper    horizontal    line),    $B(\mu^-   \to    e^-\gamma)    \sim
10^{-13}$~\cite{MEG} (lower horizontal  line), $B(\mu^- \to e^-e^-e^+)
<  10^{-12}$~\cite{PDG}, the constraints  from the  non-observation of
$\mu\to      e$      conversion      in      ${}^{48}_{22}$Ti      and
${}^{197}_{\ 79}$Au~\cite{Comment1},  $R^{\rm Ti}_{\mu e}  < 4.3\times
10^{-12}$~\cite{Titanium}  (dash-dotted)  and  $R^{\rm Au}_{\mu  e}  <
7\times   10^{-13}$~\cite{Gold}  (dash-double-dotted),   as   well  as
potential limits from a future  sensitivity to $R^{\rm Ti}_{\mu e}$ at
the $10^{-18}$ level~\cite{PRISM} (lower dash-dotted line).  Comparing
the upper with the middle  panel, one can see that $B(\mu\to e\gamma)$
becomes smaller  when the heavy sneutrino  contributions are included,
while the  other observables become larger.  The  consideration of the
mSUGRA  boundary   condition  alter   the  predictions  for   the  LFV
observables  in  non-trivial  manner,  i.e.~there are  no  regions  of
cancelation  among  terms proportional  to  ${\bf  \Omega}$ and  ${\bf
  \Omega}^2$.   Furthermore, the theoretical  predictions for  the LFV
observables may increase, especially for $\mu\to e$ conversion processes.

Figure~\ref{f2} exhibits  exclusion contours for  present experimental
limits to  LFV processes  of $\tau \to  e$ transitions:  $B(\tau^- \to
e^-\gamma)   <  3.3  \times   10^{-8}$~\cite{Babar2009}~(solid  line),
$B(\tau^- \to e^-e^-e^+) < 2.7\times 10^{-8}$~\cite{Belle2010}~(dashed
line),    and    $B(\tau^-     \to    e^-\mu^-\mu^+)    <    2.7\times
10^{-8}$\cite{Belle2010}~(dash-dotted  line).   The  dominance of  the
heavy neutrino  effects in  MSSM3NS manifests already  at $m_N\sim200$
GeV and  becomes more pronounced  than in the MSSM3Nf.   The branching
ratios for processes, such as $\tau\to 3~\mbox{leptons}$, can be $\sim
3$  times  larger than  the  one  for  $\tau\to e\gamma$  at  $m_N\sim
1000$~GeV.

Figure~\ref{f3} exhibits  exclusion contours for  present experimental
limits to LFV  processes of $\tau \to \mu$  transitions: $B(\tau^- \to
\mu^-\gamma)  <  4.4  \times  10^{-8}$~\cite{Babar2009}~(solid  line),
$B(\tau^-        \to        \mu^-\mu^-\mu^+)        <        2.1\times
10^{-8}$~\cite{Belle2010}~(dashed    line),    and    $B(\tau^-    \to
\mu^-e^-e^+)   <   1.8  \times   10^{-8}$\cite{Belle2010}~(dash-dotted
line). The exclusion contours in  all three panels are very similar to
the  corresponding  contours  for  $\tau\to e$  transitions,  but  the
dominance  of the  heavy neutrinos  is slightly  more  pronounced.  In
particular,  the heavy  neutrino  dominance in  the MSSM3NS  manifests
before  $m_N \sim 200$  GeV and  $B(\tau\to 3~\mbox{leptons})$  can be
about  5   times  larger  than  $B(\tau\to   \mu\gamma)$  at  $m_N\sim
1000$~GeV.

In  summary,  we have  shown  that  the  incorporation of  the  mSUGRA
boundary  condition  into  the  MSSM3N  leads  to  larger  theoretical
predictions  for the  LFV  observables $R_{\mu  e}$,  $\mu\to 3e$  and
$\tau\to\mbox{3  leptons}$ by  up to  a  factor of  5.  The  branching
ratios   for  the  $\ell\to\ell'\gamma$   processes  show   a  smaller
variation; they are slightly larger than those obtained in the MSSM3Nf
but smaller  than the ones  in the SM3N.  We plan to  present detailed
results of this preliminary analysis in the near future~\cite{IPP}.

\medskip
\noindent
{\bf  Acknowledgements:} 
A.I. thanks  K.  Kumeri\v  cki and L.  Popov for some  programming help.
The  work of  A.I.  was supported  by  Ministry of  Science, Sports  and
Techology  under contract  119-0982930-1016,  and the  work  of A.P.  
in part by the STFC research grant: PP/D000157/1.

\end{document}